# Probability tree algorithm for general diffusion processes

Lester Ingber[1,2] <ingber@ingber.com>, <ingber@alumni.caltech.edu>
Colleen Chen[1] <cchen@drwinvestments.com>
Radu Paul Mondescu[1] <rmondescu@drwtrading.com>
David Muzzall[1] <dmuzzall@drwinvestments.com>
Marco Renedo[1] <mrenedo@drwinvestments.com>

[1] DRW Investments, LLC, 311 S Wacker Dr, Ste 900, Chicago, IL 60606
[2] Lester Ingber Research, POB 06440 Sears Tower, Chicago, IL 60606

**ABSTRACT**

Motivated by path-integral numerical solutions of diffusion processes, PATHINT, we present a new tree algorithm, PATHTREE, which permits extremely fast accurate computation of probability distributions of a large class of general nonlinear diffusion processes.

## 1. INTRODUCTION

### 1.1. Path Integral Solution of Diffusion Processes

There are three equivalent mathematical representations of a diffusion process, provided of course that boundary and initial conditions can be properly specified in each representation. In this paper we refer to all three representations.

The Langevin rate equation for a stochastic process in $dS$ can be written as in a prepoint discretization,

$$dS = f dt + g dW ,$$

$$< dW >= 0 ,$$

$$< dW(t) dW(t') >= dt \delta(t - t') , \quad (1)$$

for general drift $f$ and standard deviation $g$ which may depend on $S$ and $t$, wherein $f$ and $g$ are understood to be evaluated at the prepoint $t$. Here, we just consider $S$ dependent, but our algorithm can easily be extended to time dependent cases and to multivariate systems.

This corresponds to a Fokker-Planck equation representing the short-time conditional probability $P$ of evolving within time $dt$,

$$\frac{\partial P}{\partial t} = - \frac{\partial (fP)}{\partial S} + \frac{1}{2} \frac{\partial^2 (g^2 P)}{\partial S^2} , \quad (2)$$

where the diffusion is given by $g^2$.

The path-integral representation for $P$ for the short-time propagator is given by

$$P(S', t'|S, t) = \frac{1}{2\pi g^2 \Delta t} \exp(-L dt)$$

$$L = \frac{(\frac{dS}{dt} - f)^2}{2g^2}$$

$$\frac{dS}{dt} = \frac{S' - S}{dt} , \; dt = t' - t . \quad (3)$$

In the above we have explicitly used the prepoint discretization [1].

### 1.2. PATHINT Motivation From Previous Study

In the above we have explicitly used the prepoint discretization, wherein $f$ and $g$ are understood to be evaluated at the prepoint $t$. In this paper, we do not require multivariate generalizations, or issues dealing with other discretizations, or explication of long-time path-integral evaluations, or issues dealing with Riemannian invariance of our distributions. There exist other references dealing with these issues in the context of calculations presented here [2-5].

Our approach is motivated by a multivariable generalization of a numerical path-integral algorithm [6-8], PATHINT, used to develop the long-time evolution of the short-time probability distribution as used in several studies in chaotic systems [9,10], neuroscience [9,11,12], and financial markets [4]. These studies suggested that we apply some aspects of this algorithm to the standard binomial tree.

### 1.3. PATHTREE Algorithms

Tree algorithms are generally derived from binomial random walks [13]. For many applications, “tree” algorithms are often used, corresponding to the above Langevin and Fokker-Planck equations [14,15]. These algorithms have typically been only well defined for specific functional forms of

$f$ and $g$.

We have previously presented a powerful PATHINT algorithm to deal with quite general $f$ and $g$ functions [4]. This general PATHTREE algorithm can be used beyond previous specific systems, affording fast reasonable resolution calculations for probability distributions of a large class of nonlinear diffusion problems.

### 1.4. Organization of Paper

Section 2 describes the standard tree algorithm. Section 3 develops our probability PATHTREE algorithm. Section 4 presents our probability calculations. Section 5 is our conclusion.

## 2. STANDARD OPTION TREE ALGORITHM

### 2.1. Binomial Tree

In a two-step binomial tree, the step up $Su$ or step down $Sd$ from a given node at $S$ is chosen to match the standard deviation of the differential process. The constraints on $u$ and $d$ are chosen as

$$ud = 1 \ , \tag{4}$$

If we assign probability $p$ to the up step $Su$, and $q = (1 - p)$ to the down step $Sd$, the matched mean and variance are

$$pSu + (1 - p)Sd = < S(t + \Delta t) > \ ,$$

$$S^2 \big( pu^2 + qd^2 - (pu + qd)^2 \big) = < \big( S(t + \Delta t) - < S(t + \Delta t) > \big)^2 > \ . \tag{5}$$

The right-hand-side can be derived from the stochastic model used.

### 2.2. Trinomial Tree

The trinomial tree can be used as a robust alternate to the binomial tree. Assume $p_u$, $p_m$ and $p_d$ are the probabilities of up jump $Su$, middle (no-)jump $S$ and down jump $Sd$, where the jumps are chosen to match the standard deviation. To match the variance of the process, the equations must satisfy

$$p_u + p_m + p_d = 1 \ ,$$

$$S(p_u u + p_m + p_d d) = < S(t + \Delta t) > \ ,$$

$$S^2 \big( p_u u^2 + p_m + p_d d^2 - (p_u u + p_m + p_d d)^2 \big) = < \big( S(t + \Delta t) - < S(t + \Delta t) > \big)^2 > \ . \tag{6}$$

## 3. PROBABILITY TREE ALGORITHM

### 3.1. General Diffusion Process

Consider the general Markov multiplicative diffusion process interpreted as an Itô prepoint discretized process, Eq. (1) with drift $f$ and diffusion $g^2$. For financial option studies the particular form of the drift $bS$ and diffusion $(\sigma S)^2$ is chosen for lognormal Black-Scholes (BS) calculations [14]. For options, the coefficient $b$ is the cost of carry, e.g., $b = r$, the risk-free rate, when $S$ is a stock price, and $b = 0$ when $S$ is a futures price [16]. The case of drift $bS$ and constant diffusion diffusion $\sigma^2$ corresponds to the Ornstein-Uhlenbeck (OU) process [17].

Our formalism is general and can be applied to other functional forms of interest with quite general nonlinear drifts and diffusions, of course provided that the region of solution does not violate any boundary or initial conditions.

Statistical properties of the $dS$ process and of any derivative one based on nonlinear transformations applied to $S$ are determined once the transition probability distribution function $P(S, t|S_0, t_0)$ is known, where the 0 index denotes initial values of time and of the stochastic variable $S$. Transformation are common and convenient for BS,

$$z = \ln S \ , \tag{7}$$

yielding a simple Gaussian distribution in $z$, greatly simplifying analytic and numerical calculations.

The probability distribution can be obtained by solving the associated forward Fokker-Planck equation Eq. (2). Appropriate boundaries and initial condition must be specified, e.g., $P(S|S_0) = \delta(S - S_0)$.

In general cases, the Fokker-Planck equation is rather difficult to solve, although a vast body of work is devoted to it [17]. The particular BS and OU cases possess exact results.

Our goal is to obtain the solution of Eq. (1) for the more general process. A quite general code, PATHINT [4], works fine, but it is much slower than the PATHTREE algorithm we present here.

### 3.2. Deficiency of Standard Algorithm to Order $\sqrt{dt}$

We briefly describe the CRR construction of the binomial tree approximation [18].

A tree is constructed that represents the time evolution of the stochastic variable $S$. $S$ is assumed to take only 2 values, $u$, (up value), and $d$ (down value) at moment $t$, given the value $S$ at moment $t - \Delta t$. The probabilities for the up and down movements are $p$ and $q$, respectively. The 4 unknowns $\{u, d, p, q\}$ are calculated by imposing the normalization of the probability and matching the first two moments conditioned by the value $S$ at $t - \Delta t$, using the variance of the exact probability distribution $P(S, t|S_0, t_0)$. One additional condition is arbitrary and is usually used to symmetrize the tree, e.g., $ud = 1$.

The main problem is that the above procedure cannot be applied to a general nonlinear diffusion process as considered in Eq. (1), as the algorithm involves a previous knowledge of terms of $O(\Delta t)$ in the formulas of quantities $\{u, p\}$ obtained from a finite time expansion of the exact solution $P$ sought. Otherwise, the discrete numerical approximation obtained does not converge to the proper solution.

This observation can be checked analytically in the BS CRR case by replacing the relation $u = \exp(\sigma \Delta t)$ [15] with $u = 1 + \sigma\sqrt{\Delta t}$, and deriving the continuous limit of the tree. This also can be checked numerically, as when $\{u, p\}$ are expanded to $O(\Delta t)$, the proper solution is obtained.

### 3.3. Probability PATHTREE

As mentioned previously, a general path-integral solution of the Fokker-Plank equation, including the Black-Scholes equation, can be numerically calculated using the PATHINT algorithm. Although this approach leads to very accurate results, it is computationally intensive.

In order to obtain tree variables valid up to $O(\Delta t)$, we turn to the short-time path-integral representation of the solution of the Fokker-Planck equation, which is just the multiplicative Gaussian-Markovian distribution [1,19]. In the prepoint discretization relevant to the construction of a tree,

$$P(S', t'|S, t) = \frac{1}{\sqrt{2\pi \Delta t g^2}} \exp\left( -\frac{(S' - S - f dt)^2}{2 g^2 \Delta t} \right)$$

$$\Delta t = t' - t \tag{8}$$

valid for displacements $S'$ from $S$ “reasonable” as measured by the standard deviation $g\sqrt{\Delta t}$, which is the basis for the construction of meshes in the PATHINT algorithm.

The crucial aspects of this approach are: There is no a priori need of the first moments of the exact long-time probability distribution $P$, as the necessary statistical information to the correct order in time is contained in the short-time propagator. The mesh in S at every time step need not recombine in the sense that the prepoint-postpoint relationship be the same among neighboring $S$ nodes, as the short-time probability density gives the correct result up to order $O(\Delta t)$ for any final point $S'$. Instead, we use the natural metric of the space to first lay down our mesh, then dynamically calculate the evolving local short-time distributions on this mesh.

We construct an additive PATHTREE, starting with the initial value $S_0$, with successive increments

$$S_{i+1} = S_i + g\sqrt{\Delta t} \ , \ S_i > S_0$$

$$S_{i-1} = S_i - g\sqrt{\Delta t}\ , \ S_i < S_0\ , \tag{9}$$

where $g$ is evaluated at $S_i$. We define the up and down probabilities $p$ and $q$, resp., in an abbreviated notation, as

$$p = \frac{P(i+1|i;\Delta t)}{P(i+1|i;\Delta t) + P(i-1|i;\Delta t)}$$

$$q = 1 - p\ . \tag{10}$$

where the short-time transition probability densities $P$'s are calculated from Eq. (8). Note that in the limit of small $\Delta t$,

$$\lim_{\Delta t \to 0} p = \frac{1}{2}\ . \tag{11}$$

### 3.3.1. Continuous Limit of the PATHTREE Algorithm

In either the upper or lower branches of the tree ($S_i > S_0$ or $S_i < S_0$, resp.), we always have the postpoint $S_{i\pm1}$ in terms of the prepoint $S_i$, but we also need the inverses to find the asymptotic $\lim_{\Delta \to 0}$ of our PATHTREE building technique. For example, for the upper branch case,

$$S_{i-1} \approx S_i - g\sqrt{\Delta t} + (g\,\frac{\partial g}{\partial S})\Delta t + O(\Delta t^{3/2})\ . \tag{12}$$

This expression must be used to extract the down change $d$ ($S_{i-1} = S_i + d$), for comparison to the standard tree algorithm.

The continuous limit of the previous tree building procedure is obtained by Taylor expanding all factors up to terms $O(\Delta t)$ and as functions of the prepoint $S_i$ [15]. This leads to

$$\frac{pu + qd}{\Delta t} \approx f + O(\Delta t^{1/2})$$

$$\frac{pu^2 + qd^2}{\Delta t} \approx g^2 + O(\Delta t^{1/2})\ , \tag{13}$$

from which the correct partial differential equation, Eq. (2), is recovered up to $O(\Delta t)$.

In implementing the PATHTREE algorithm, good numerical results are obtained in the parameter region defined by the convergence condition

$$\left( \left| g\,\frac{\partial g}{\partial S_i} \right| dt + g\sqrt{dt} \right) / S_i \ll 1\ . \tag{14}$$

This insures the proper construction of the tree to order $O(\Delta t)$.

### 3.3.2. Treating Binomial Oscillations

Binomial trees exhibit by construction a systematic oscillatory behavior as a function of the number of steps in the tree (equivalently, the number of time slices used to build the tree), and the new building algorithm based on the short-time propagator of the path-integral representation of the solution of the Fokker-Planck equation has this same problem. A common practice [20] is to perform averages of runs with consecutive numbers of steps, e.g.,

$$C = \frac{C_{N+1} + C_N}{2}\ , \tag{15}$$

where $C_N$ signifies the value calculated with $N$ number of steps.

#### 3.3.3. Inappropriate Trinomial Tree

Another type of tree is the trinomial tree discussed above, equivalent to the explicit finite difference method [14,15]. If we were to apply this approach to our new PATHTREE algorithm, we would allow the stochastic variable $S$ to remain unchanged after time step $\Delta t$ with a certain probability $p_m$. However, in our construction the trinomial tree approach is not correct, as the deterministic path is dominant in the construction of the probabilities $\{p_u, p_m, p_d\}$, and we would obtain

$$\lim_{\Delta t \to 0} p_u = p_d = 0 \ ,$$

$$\lim_{\Delta t \to 0} p_m = 1 \ . \quad (16)$$

### 3.4. Linear and Quadratic Aspects of Numerical Implementation

PATHTREE computes the expected value of a random variable at a later time given the diffusion process and an initial point. The algorithm is similar to a binomial tree computation and it consists of three main steps: computation of a mesh of points, computation of transition probabilities at those points and computation of the expected value of the random variable.

The first step is the creation of a one dimensional mesh of points with gaps determined by the second moment of the short term distribution of the process. The mesh is created sequentially, starting from the initial point, by progressively adding to the last point already determined (for the upward part of the mesh) the value of the standard deviation of the short term distribution with the same point as prepoint. In a similar fashion we create the mesh downwards, this time by subtracting the standard deviations. The procedure takes a linear amount of time on the number of time slices being considered and contributes very little to the overall time of the algorithm.

In the second step an array of up and down probabilities is created. These probabilities are the values of the short term transition probability density function obtained by using the current point as prepoint and the two neighboring points as post points. The probabilities are renormalized to sum to unity. This procedure takes a linear amount of time on the number of time slices. Notice that the probabilities only depend on the current point and not on time slice, hence only two probabilities are computed per element of the array of points.

The third step is the computation of the expected value of the random variable. For example, the option price $C$ is developed by marching backwards along the tree and applying the risk-neutral evaluation

$$C(S_i, t - \Delta t) = e^{-r\Delta t}[pC(S_{i+1}, t) + qC(S_{i-1}, t)] \ . \quad (17)$$

We emphasize again that in Itô terms the prepoint value is $S_i$. This part works as a normal binomial tree algorithm. The algorithm uses the expected values at one time slice to compute the expected values at the previous one. The bulk of the time is spent in this part of the algorithm because the number of iterations is quadratic on the amount of time slices. We managed to optimize this part by reducing each iteration to about 10 double precision operations.

In essence, this algorithm is not slower than standard binomial trees and it is very simple to implement.

## 4. CALCULATION OF PROBABILITY

### 4.1. Direct Calculation of Probability

We can calculate the probability density function by first recursively computing the probabilities of reaching each node of the tree. This can be performed efficiently thanks to the Markov property. To compute the density function we need to rescale these probabilities by the distance to the neighboring nodes: the more spread the nodes are, the lower the density. We can estimate the probability density as follows: First we compute the probability of reaching each final node of the tree. We do this incrementally by first computing the probabilities of reaching nodes in time slice 1, then time slice 2 and so forth. At time slice 0, we know that the middle node has probability 1 of being reached and all the others have probability 0. We compute the probability of reaching a node as a sum of two contributions from the previous time slice. We reach the node with transition *pu* from the node below at the previous slice, and with transition *pd* from the node above. Each contribution is the product of the probability at the previous node times the transition to the current node. This formula is just a discretized version of the Chapman-Kolmogorov equation

$$p(x_j, t_i) = p(x_{j-1}, t_{i-1}) pu_{j-1} + p(x_{j+1}, t_{i-1}) pd_{j+1} \ . \tag{18}$$

Now that we have computed the absolute probabilities at the final nodes, we can give a proper prepoint-discretized estimation of the density by scaling the probabilities by the spread of the S values. For the upper half of the tree we divide the probability of each final node by the size of the lower adjacent interval in the mesh: $\text{density}_i = p_i/(S_i - S_{i-2})$. (Note: We use index $S_{i-2}$ because the binomial tree is constructed over a trinomial tree. In this way we can keep in memory all the nodes but only half of the nodes though are true final nodes.) If there is a final middle node we divide its probability by the average of sizes of the two adjacent intervals, that is: $\text{density}_i = p_i/((S_{i+2} - S_{i-2})/2)$. For the lower half of the mesh we divide the probability by the upper adjacent gap in the mesh: $\text{density}_i = p_i/(S_{i+2} - S_i)$.

### 4.2. Numerical Derivatives of Expectation of Probability

The probability $P$ can be calculated as a numerical derivative with respect to strike $X$ of a European Call option, taking the risk-free rate $r$ to be zero, given an underlying $S_0$ evaluated at time $t = 0$, with strike $X$, and other variables such as volatility $\sigma$, cost of carry $b$, and time to expiration $T$ suppressed here for clarity, $C(S_0, 0; X)$,

$$P[S(T)|S(t_0)] \underset{S(T)\equiv X}{=} P[X|S(t_0)] = \frac{\partial^2 C}{\partial X^2} \tag{19}$$

This calculation of the probability distribution is dependent on the same conditions necessary for any tree algorithm, i.e., that enough nodes are processed to ensure that the resultant evaluations are a good representation of the corresponding Fokker-Planck equation, addressed above, and that the number of iterations in PATHTREE are sufficient for convergence.

#### 4.2.1. Alternative First Derivative Calculation of Probability

An alternative method of calculating the probability $P$ a a first-order numerical derivative, instead of as second-order derivative, with respect to $X$ is to define a function $C_H$ using the Heaviside step-function $H(S, X) = 1$ if $S \ge X$ and 0 otherwise, instead of the Max function at the time to expiration. This yields

$$P[S(T)|S(t_0)] \underset{S(T)\equiv X}{=} P[X|S(t_0)] = -\frac{\partial C_H}{\partial X} \tag{20}$$

Sometimes this is numerically useful for sharply peaked distributions at the time of expiration, but we have found the second derivative algorithm above to work fine with a sufficient number of epochs.

Our tests verify that the three methods above give the same density. We consider the numerical-derivative calculations a very necessary baseline to determine the number of epochs required to get reasonable accuracy.

#### 4.2.2. Oscillatory Corrections

Fig. 1 illustrates the importance of including oscillatory corrections in any binomial tree algorithm. When these are included, it is easy to see the good agreement of the BS PATHTREE and OU PATHTREE models.

### 4.3. Comparison to Exact Solutions

Fig. 2 gives the calculated probability distribution for the BS and OU models, compared to their exact analytic solutions.

## 5. CONCLUSION

We have developed a path-integral based binomial PATHTREE algorithm that can be used in a variety of stochastic models. This algorithm is simple, fast and can be applied to diffusion processes with quite arbitrarily nonlinear drifts and diffusions.

As expected, this PATHTREE algorithm is not as strong as PATHINT [4], as PATHINT can include details of an extremely high dimensional tree with complex boundary conditions.

For PATHINT, the time and space variables are determined independently. I.e., the ranges of the space variables are best determined by first determining the reasonable spread of the distribution at the final time epoch. For PATHTREE just one parameter, the number of epochs $N$, determines the mesh for both time and the space variables. This typically leads to a growth of the tree, proportional to $\sqrt{N}$, much faster than the spread of the distribution, so that much of the calculation is not relevant.

However, this PATHTREE algorithm is surprisingly robust and accurate. Similar to PATHINT, we expect its accuracy to be best for moderate-noise systems.

## ACKNOWLEDGMENTS

We thank Donald Wilson for his financial support.

**FIGURE CAPTIONS**

FIG. 1. The oscillatory correction, an average of $N$ and $N + 1$ iteration solutions, provides a simple and effective fix to the well-known oscillations inherent to binomial trees. The uncorrected Black-Scholes binomial tree (a) can be compared to the Black-Scholes tree with oscillatory correction (b). In (c), the Ornstein-Uhlenbeck binomial tree also be robustly corrected as shown in (d). The BS PATHTREE model shown in (e) can be compared to the Black-Scholes case shown in (b). The OU PATHTREE model (f) is equivalent to the Ornstein-Uhlenbeck model in (d). Parameters used in these calculations are: $S$ = 50.0, $X$ = 55.0, $T$ = 1.0, $r$ = 0.0675, $b$ = 0, $\sigma$ = 0.20, and $N$ = 300.

FIG. 2. Probability distributions for the PATHTREE binomial model as described in the text. In (a), bar graphs indicate OU PATHTREE agrees well with the exact Ornstein-Uhlenbeck distribution shown in the black line. In (b), the bar graphs indicate BS PATHTREE agrees well with the exact Black-Scholes distribution shown in the black line. Parameters are the same as in Fig. 1.

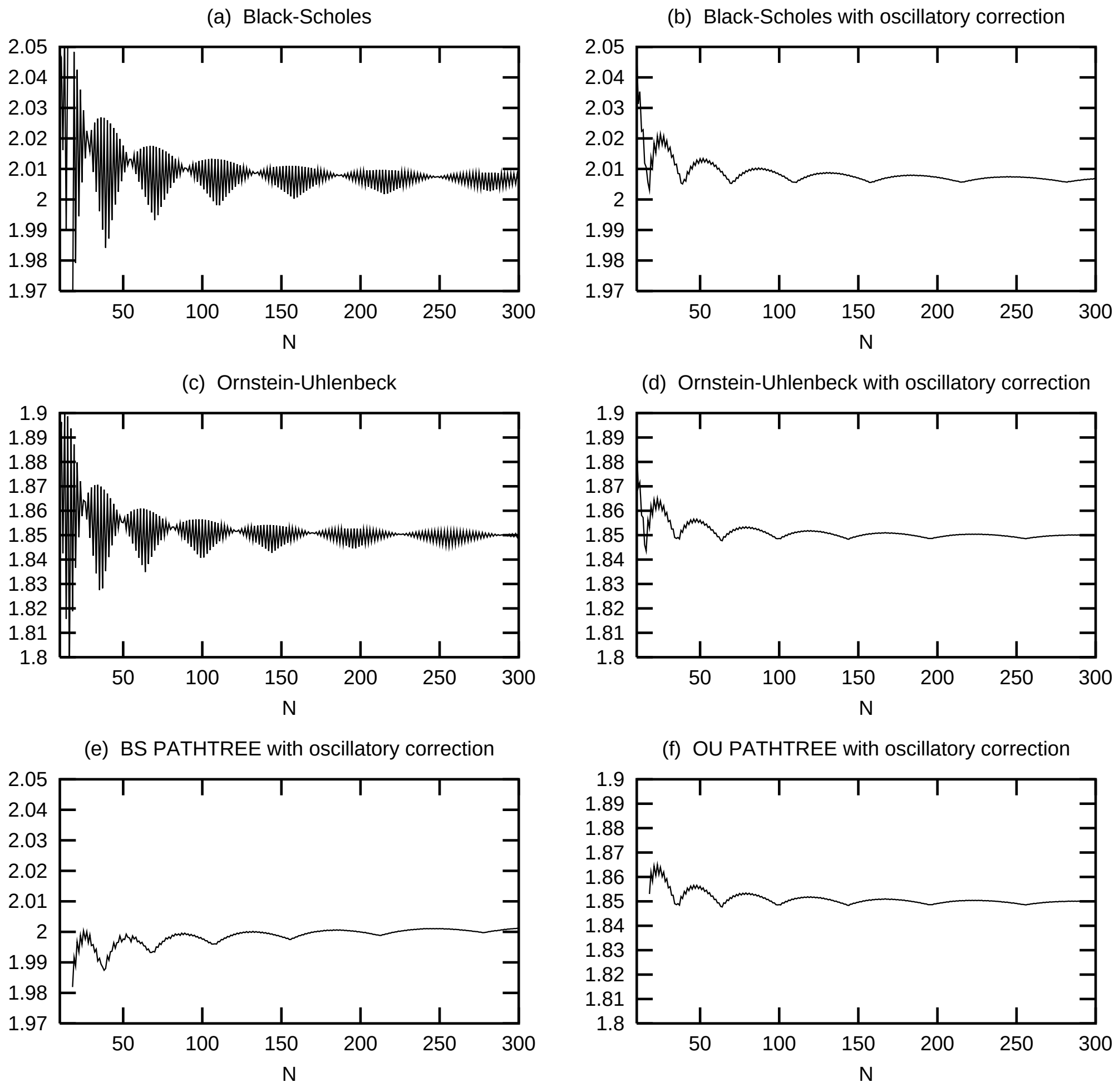

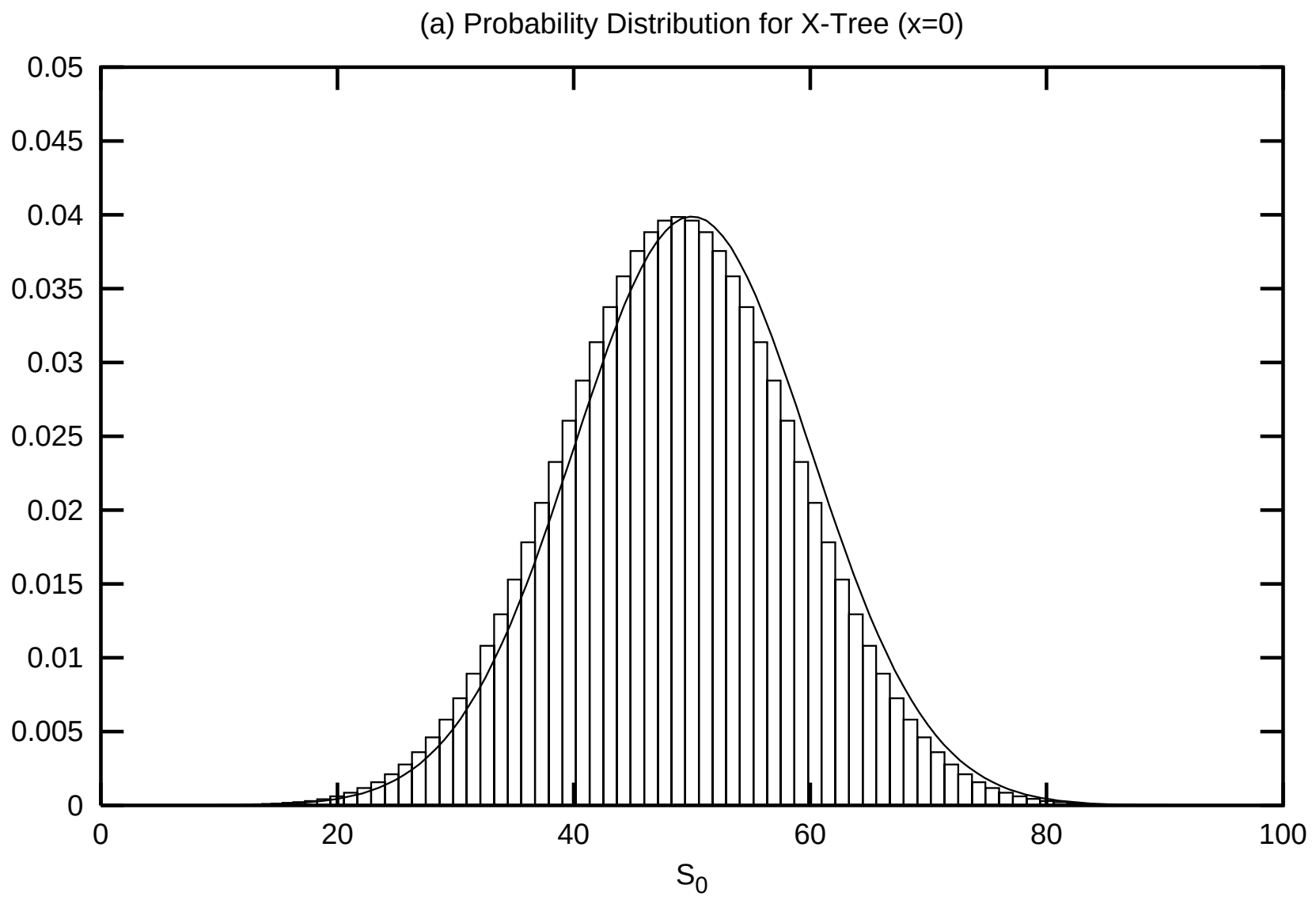

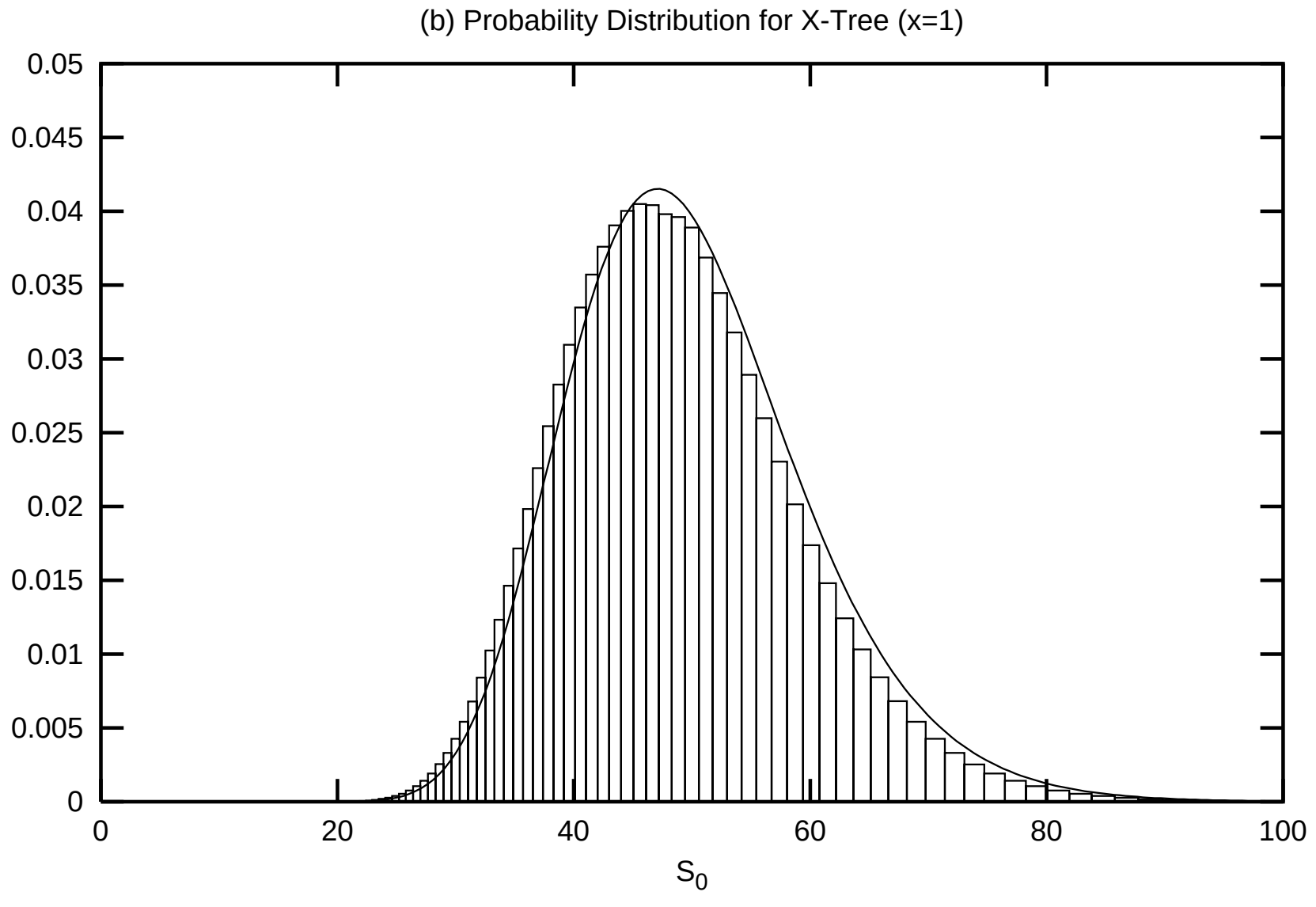